\documentstyle[12pt]{article}
\renewcommand
\baselinestretch{1.3}

\begin{document}

\def\beq{\begin{equation}}
\def\eeq{\end{equation}}
\def\bce{\begin{center}}
\def\ece{\end{center}}
\def\bea{\begin{eqnarray}}
\def\eea{\end{eqnarray}}
\def\ben{\begin{enumerate}}
\def\een{\end{enumerate}}
\def\ul{\underline}
\def\ni{\noindent}
\def\nn{\nonumber}
\def\bs{\bigskip}
\def\ms{\medskip}
\def\wt{\widetilde}
\def\brr{\begin{array}}
\def\err{\end{array}}

%\hfill HUPD-93?

\hfill January 1994

\vspace*{3mm}

\begin{center}

{\LARGE \bf
Chiral symmetry breaking in the $d=3$ Nambu-Jona-Lasinio model in
curved
spacetime}

\vspace{4mm}

\renewcommand
\baselinestretch{0.8}
\medskip

{\sc E. Elizalde}
\footnote{E-mail: eli@ebubecm1.bitnet, eli@zeta.ecm.ub.es} \\
Department E.C.M. and I.F.A.E., Faculty of Physics,
University of  Barcelona, \\ Diagonal 647, 08028 Barcelona, \\
and Center for Advanced Studies, C.S.I.C., Cam¡ de Santa B…rbara,
\\
17300 Blanes, Catalonia, Spain \\
{\sc S.D. Odintsov}
\footnote{E-mail: odintsov@ebubecm1.bitnet; on leave from
Tomsk Pedagogical Institute, 634041 Tomsk, Russian Federation.} \\
Department E.C.M., Faculty of Physics,
University of  Barcelona, \\  Diagonal 647, 08028 Barcelona,
Catalonia,
Spain \\
and \
{\sc Yu.I. Shil'nov}\\
Department of Theoretical Physics, Kharkov State University, \\
Svobody sq. 4, Kharkov 310077, Ukraine

\renewcommand
\baselinestretch{1.4}

\vspace{5mm}

{\bf Abstract}

\end{center}

The phase structure of the $d=3$ Nambu-Jona-Lasinio model in curved
spacetime is considered to leading order in the $1/N$--expansion
and in
the linear curvature approximation. The possibility of a
curvature-induced first-order phase transition is investigated
numerically. The dynamically generated fermionic mass is calculated
for
some values of the curvature.

\vspace{4mm}

\newpage

\ni 1. \ Four-fermion models provide a very useful laboratory for
the analytical study of composite bound states and dynamical
symmetry
breaking. A particularly interesting theory belonging to this class
is
the $d=3$ four-fermion model, which is the natural $d=3$ analog of
the
well-known Nambu-Jona-Lasinio model \cite{1} (for a recent
discussion, in connection with dynamical symmetry breaking in
electroweak interactions, see \cite{2}). This model is known to be
renormalizable
in the $1/N$--expansion \cite{3}, and has a quite interesting phase
structure
in  $d=3$ flat space with non-trivial topology \cite{4}. From a
different point of view, the $d=3$ four-fermion model can be
considered
as the high-temperature phase of the corresponding $d=4$ model,
what
gives an additional motivation for its study.
Let us also recall that four-fermion models are interesting, in
general,
as effective theories of QCD (for a comprehensible introduction to
QCD, see \cite{10}).

Recently, a careful investigation of the $d=4$ four-fermion theory
(i.e., the Nambu-Jona-Lasinio model \cite{1}) in curved spacetime
has
been undertaken \cite{5}-\cite{7}. In particular, it has been shown
in
detail the existence of a first-order phase transition ---induced
by the
curvature--- from the chiral symmetric phase to the chiral
non-symmetric
one \cite{6,7}. The dynamical generation of the fermion mass due to
curvature effects has also been discussed ---a process that may
have
interesting applications in the very early universe (see, for
example,
\cite{8}).

In the present note we will investigate the $d=3$ four-fermion
model in
curved spacetime. The effective potential for the composite field
$\bar{\psi} \psi$ will be calculated in the $1/N$--expansion and
its
phase structure will be discussed. In particular, the possibility
of
chiral symmetry
breaking and the influence of curvature in this phenomenon will be
studied.

 \bs

\ni 2. \ Our starting point is the Lagrangian of
the $d=3$ four-fermion model in curved spacetime:
\beq
L= \bar{\psi}^i \gamma^\mu (x) \nabla_\mu \psi^i -
\frac{N\sigma^2}{2\lambda} - \sigma \bar{\psi}^i \psi_i,
\label{1}
\eeq
where $i=1, \ldots, N$, and $N$ is the number of fermion species.
The
Lagrangian (\ref{1}) is equivalent to the four-fermion model, what
is
easily seen by solving for the equation which involves the
composite
field $\sigma$. Working in the linear curvature approximation
and using the technique described in detail in \cite{6} ---for the
$d=4$
Nambu-Jona-Lasinio model--- we obtain, after some calculations, the
effective potential for the composite field $\sigma$ to leading
order in the $1/N$--expansion (we shall spare the reader the
details of the rather tedious computation)
\bea
V(\sigma) &=& \frac{\sigma^2}{2\lambda} - \frac{1}{3\pi^2} \left[
\Lambda^3 \ln \left( 1 + \frac{\sigma^2}{\Lambda^2} \right) +
2\sigma^2 \left( \Lambda - \sigma \arctan \frac{\Lambda}{\sigma}
\right) \right] \nn \\
&&-  \frac{R}{36\pi^2}  \left(
\frac{2\sigma^2\Lambda}{\Lambda^2+\sigma^2} -3\sigma  \arctan
\frac{\Lambda}{\sigma} \right).
\label{2}
\eea
Here $\Lambda$ is a cut-off, and the curvature is considered to be
arbitrary, but small, so that it is meaningful to keep only linear
curvature corrections to the flat potential, dropping ${\cal O}
(R^2)$-terms and those of higher order.

We now start with the analysis of Eq. (\ref{2}). Consider
first the case $R=0$. Then, if $\lambda < \lambda_c =
\pi^2/(2\Lambda)$, it turns out that $\sigma =0$ is the minimum of the
potential
and the theory is in its chiral symmetric phase. If, on the other
hand, $\lambda > \lambda_c = \pi^2/(2\Lambda)$, then $\sigma =0$ is
a maximum, and the minimum of the effective potential is now given
by the solution of the equation
\beq
\frac{\sigma}{\Lambda} \arctan \frac{\Lambda}{\sigma}  =1-
\frac{\lambda_c}{\lambda}.
\label{3}
\eeq
In this region the chiral symmetry of the classical Lagrangian is
broken.
One may ask for finiteness of $\sigma$ as $\Lambda \rightarrow
\infty$.
This can be obtained by means of the corresponding standard
renormalization of $\lambda$:
\beq \lambda = \frac{\lambda_c \ \frac{\Lambda}{\sigma} }{
\frac{\Lambda}{\sigma}- \arctan \frac{\Lambda}{\sigma} }.
\eeq
The corresponding ``$\beta$-function" is defined as (see, for
instance, Ref. \cite{9})
\beq
\beta = \lim_{\Lambda \rightarrow \infty} \frac{\partial
\lambda}{\partial \log \Lambda} = \frac{\lambda}{\lambda_c} \left(
\lambda_c -\lambda \right).
\eeq
Such a $\beta$-function indicates that $\lambda_c$ is an
ultraviolet stable fixed point.
\bs

\ni 3. \ Let us now proceed with the numerical computation of the
effects of curvature on chiral symmetry breaking. The conditions
for a first-order phase transition are given by
 \beq
\left.  V(\sigma) \right|_{\sigma = \sigma_0} =0, \ \ \ \ \ \
\left. \frac{\partial V(\sigma)}{\partial \sigma} \right|_{\sigma
=   \sigma_0} =0,
\eeq
which yield a minimum of $V(\sigma)$.
They can be written as: $y(t)=0$, $y'(t)=0$, namely
\bea
 \frac{t^2}{x}  - \frac{1}{3} \left[
 \ln \left( 1 + t^2 \right) +
2t^2 \left( 1- t \arctan \frac{1}{t}
\right) \right] -  \frac{\rho}{36}  \left(
\frac{2t^2}{1+t^2} -3t  \arctan \frac{1}{t} \right) &=& 0, \nn \\
 \frac{2t}{x}  - 2t \left( 1- t \arctan \frac{1}{t}
\right) -  \frac{\rho}{36}  \left[
\frac{4t}{(1+t^2)^2}+ \frac{3t}{1+t^2} -3 \arctan \frac{1}{t}
\right] &=& 0
\eea
where
\beq
t \equiv \frac{\sigma}{\Lambda}, \ \ \ \ y \equiv
\frac{V}{\Lambda^3}, \ \ \ \ x \equiv \frac{\lambda}{\lambda_c},
 \ \ \ \ \rho \equiv \frac{R}{\Lambda^2}.
\eeq
It can be checked that the solution, $t_0$, of these two
equations readily satisfies $y''(t_0) > 0$ (for $x>1$), with
\beq
y''(t) =\frac{2}{\pi^2} \left\{ \frac{1}{x}  - 2+
\frac{1}{1+t^2} + 2t \arctan \frac{1}{t} -  \frac{\rho}{36}
\left[ \frac{8}{(1+t^2)^3}- \frac{3}{(1+t^2)^2}
\right] \right\},
\eeq
and it is thus, in fact, a minimum of $y(t)$.
Numerical  results, corresponding to several values of
$\lambda /\lambda_c$, together with the respective values of the
critical
curvature $R_c /\Lambda^2$ in each case, are given in Table 1.
\begin{table}

\begin{center}

\begin{tabular}{|c|c|c|}
\hline \hline
 $\lambda/\lambda_c$  & $R_c /\Lambda^2$ & $\sigma_0 /\Lambda$ \\
\hline\hline  1.03 & 0.00158 & 0.014 \\
 \hline 1.10 & 0.016 & 0.046 \\
 \hline 1.27 & 0.10 & 0.12 \\
 \hline 2.00 & 0.80 & 0.39 \\
\hline \hline \end{tabular}

 \caption{{\protect\small Numerical values of the
dynamical fermionic mass $\sigma_0 /\Lambda$ at the point of the
curvature-induced phase transition
 ---together with the respective values of the critical
curvature $R_c /\Lambda^2$---
corresponding to a sample of values of $x \equiv \lambda /\lambda_c
>1$.}}

\end{center}

\end{table}

Table 1 shows how the dynamically generated fermionic mass changes
in
terms of the value of the corresponding critical curvature, for
several
particular values of $\lambda > \lambda_c$. Notice that in all the
cases
considered, the critical curvature is always smaller than 1, and
also
clearly  $R_c /\Lambda^2 < \sigma_0 /\Lambda$ (this improves for
$\lambda$
approaching $\lambda_c$), while $R_c$ is of the order of
$\sigma_0^2$.
However, observe also that while this last bound improves when
$\lambda$ departs
from $\lambda_c$, $\lambda > \lambda_c$, then the value of $R_c
/\Lambda^2$ approaches 1. One reaches a compromise for
\[
1.1 < \frac{\lambda}{\lambda_c} < 1.5,
\]
where all the requirements of the theory are met. The results are
given
in Figs. 1-3, where the values of $V/\Lambda^3$ as a function of
$\sigma/\Lambda$ are depicted, for $x \equiv \lambda/\lambda_c
=1.272$
and $R_c /\Lambda^2$ equal to $0.05$ (Fig. 1), $0.10$ (Fig. 2), and
$0.15$ (Fig. 3), respectively. Finally, Fig. 4 is a
surface plot of the effective potential $V/\Lambda^3$ as a function
of
both
$\sigma /\Lambda$ and $R /\Lambda^2$,  in the intervals $[0,0.2]$
and
$[0.05,0.15]$, respectively,  for fixed  $x \equiv
\lambda/\lambda_c
=1.272$. Notice that Figs. 1-3 correspond to the extreme and middle
$xz$-sections of this surface plot.
\bs

\ni 4. \ To summarize, we have here investigated the influence of
curvature to the chiral symmetry breaking pattern of the $d=3$
Nambu-Jona-Lasinio model. As  happens in the case of the
four-dimensional model \cite{6} (and within the approximation at
which we are working), we have shown that a first-order phase
transition induced by curvature does also take place for $d=3$. Of
course, this is a result obtained by using the linear curvature
approximation. It would be of interest to study the same model in
the strong curvature phase, for instance, considering it immersed
in a general De Sitter space. Then, the phase structure at
arbitrarily high curvature could be investigated. There exists thus
the possibility that chiral symmetry could be broken even in the
region $\lambda < \lambda_c$ ---where the corresponding theory in
flat space is in its chiral symmetric phase.

\vspace{5mm}

\noindent{\large \bf Acknowledgments}

SDO would like to thank T. Muta and T. Inagaki for helpful
discussions, and the members of the Dept. ECM, Barcelona
University, for their kind hospitality.
This work has been  supported by DGICYT (Spain) and by CIRIT
(Generalitat de Catalunya).

\newpage

\noindent{\Large \bf Figure captions}
\bigskip

\noindent{\bf Figure 1}.
Plot of the effective potential $V/\Lambda^3$ as a function of
$\sigma /\Lambda$ in the interval $[0,0.2]$,
for fixed values of  $x \equiv \lambda/\lambda_c =1.272$
and $R /\Lambda^2 = 0.05$.
\bigskip

\noindent{\bf Figure 2}.
The effective potential $V/\Lambda^3$ as a function of
$\sigma /\Lambda$ in the interval $[0,0.2]$,
for the same  $x \equiv \lambda/\lambda_c =1.272$, but now
$R /\Lambda^2 \equiv
R_c /\Lambda^2 = 0.10$.
The numerical value of the dynamically generated fermionic mass
$\sigma_0 /\Lambda =0.12$ can be explicitly taken from the plot in
this case.
\bigskip

\noindent{\bf Figure 3}.
Plot of $V/\Lambda^3$ as a function of
$\sigma /\Lambda$ in the interval $[0,0.2]$,
again for  $x \equiv \lambda/\lambda_c =1.272$, but now
$R /\Lambda^2 = 0.15$.
\bs

\noindent{\bf Figure 4}.
Surface plot of $V/\Lambda^3$ as a function of both
$\sigma /\Lambda$ and $R /\Lambda^2$,  in the intervals $[0,0.2]$
and
$[0.05,0.15]$, respectively,  for fixed  $x \equiv
\lambda/\lambda_c
=1.272$. Figs. 1-3 correspond to the extreme and middle $xz$-sections
of this plot.

\end{document}